\documentclass[twocolumn]{aastex62}
\usepackage{graphicx}

\usepackage{lineno}

\usepackage{color}
\usepackage{epstopdf}
\usepackage{amsmath}
\usepackage{appendix}
\usepackage{bm}
\usepackage{soul}

\newcommand{\msun}{\textup{M}_\odot}
\newcommand{\idotI}{ ( \hat{\bf i} \cdot \hat{\bf I} ) }

\begin{document}

\title{ A Stability Timescale for Non-Hierarchical Three-Body Systems }
\shorttitle{ Stability of Non-Hierarchical Triples }
\shortauthors{Zhang, Naoz, and Will}

\author[0000-0002-7611-8377]{Eric Zhang}
\affiliation{Department of Physics and Astronomy, University of California, Los Angeles, CA 90095, USA}
\affiliation{Mani L. Bhaumik Institute for Theoretical Physics, Department of Physics and Astronomy, UCLA, Los Angeles, CA 90095, USA}
\affiliation{Department of Physics and Astronomy, University of California, Riverside, CA, 92507, USA}

\author[0000-0002-9802-9279]{Smadar Naoz}
\affiliation{Department of Physics and Astronomy, University of California, Los Angeles, CA 90095, USA}
\affiliation{Mani L. Bhaumik Institute for Theoretical Physics, Department of Physics and Astronomy, UCLA, Los Angeles, CA 90095, USA}

\author[0000-0001-8209-0393]{Clifford M. Will}
\affiliation{Department of Physics, University of Florida, Gainesville, FL 32611, USA}
\affiliation{Institut d'Astrophysique, Sorbonne Universit\'e, 75014 Paris, France}

\begin{abstract}
The gravitational three-body problem is a fundamental problem in physics and has significant applications to astronomy. Three-body configurations are often considered stable as long the system is hierarchical; that is, the two orbital distances are well-separated. However, instability, which is often associated with significant energy exchange between orbits, takes time to develop. Assuming two massive objects in a circular orbit and a test particle in an eccentric orbit, we develop an analytical formula estimating the time it takes for the test particle's orbital energy to change by an order of itself.  We show its consistency with results from N-body simulations. For eccentric orbits in particular, the instability is primarily driven not by close encounters of the test particle with one of the other bodies, but by the fundamental susceptibility of eccentric orbits to exchange energy  at their periapsis.
Motivated by recent suggestions that the galactic center may host an intermediate-mass black hole (IMBH) as a companion to the massive black hole Sgr A*, we use our timescale to explore the parameter space that could harbor an IMBH for the lifetime of the S-cluster of stars surrounding Sgr A*. Furthermore, we show that the orbit of an S-star can be stable for long timescales in the presence of other orbital crossing stars, thus suggesting that the S-cluster may be stable for the 
lifetimes of its member stars.

\end{abstract}


\section{Introduction}\label{sec:Intro}

The stability of triple-body systems is a pervasive problem in astrophysics. The three-body problem describes the dynamics of systems ranging from planetary systems to the orbits of stars and compact objects. A ``stable'' bound three-body system is loosely defined as one in which the energy of each orbit stays roughly the same, and ``instability'' is associated with systems in which the two orbits exchange energy. As a dramatic example, a bound orbit in a three-body system becomes unbound if its energy changes from negative to positive, and such systems are said to be unstable. Another example for instability can be considered an exchange process between the two orbits, which also dramatically changes the energy of each orbit. 

The general approach in the literature often relies on a criterion by which the system could be considered stable “in the long run”. A common approach to developing such criteria is to designate hierarchical systems, or those in which the two orbital distances are well-separated, as long-term stable.
{The question of stability is then equivalent to determining a critical distance between the two orbits, where the system switches from hierarchical to non-hierarchical.}

The most straightforward such criterion involves the Hills mechanism \citep{Hills88}. For a system composed of a binary and a tertiary, the Hills critical distance is the separation between the tertiary and the inner binary at which the tertiary’s gravitational potential exceeds the primary’s gravitational potential. In this case, the secondary body in the inner binary may hop between the primary and the tertiary, yielding a new configuration of a tertiary-secondary and a primary. In the co-planar case, the functional expression of this place represents the first Lagrange point, which indicates the position where the gradient of the potential in the rotating frame is zero \citep[e.g.,][]{Murray+00,Binney+TremaineBook}.
{{
A stability criterion may be obtained by requiring that the tertiary is never closer to either of the other bodies than the Hills critical distance. The criterion has been improved by \citet{Hamilton+91} and \citet{Grishin+17}.
}}

The stability of three massive objects has a generalized, {{hierarchy-based,}} stability condition often used in the literature \citep[e.g.,][]{Mardling+01}, 
{{
and similar stable-unstable boundaries were derived by \citet{Eggleton+95, Petrovich15, Tory+22, Vynatheya+22, Hayashi+22}.}} Considering hierarchical systems, where one mass orbits on a tight configuration about the primary and a tertiary is on a wider orbit, a {{condition is often used}} to estimate the long-term stability {{against high-eccentricity excitations due to secular dynamics \citep[e.g.,][]{Ivanov+05,LN11,Katz+12,NaozSilk14,Antonini+14,Bode+14} and non-secular perturbations to secular dynamics \citep[e.g.,][]{Antognini+14, Luo+16, Grishin+18QuasiSec, Bhaskar+21}.}}

However, instability is a time-sensitive concept, and not every non-hierarchical system is instantaneously unstable. {{
\citet{Myllari+18} noted the dependence of stability on time, albeit without an explicit timescale. Additionally, \citet{Mushkin+20} developed a stability timescale for the outer orbit in hierarchical systems, based on formulae for secular energy exchange \citep[e.g.,][]{RoyHaddow03}. Further studies on the time-dependence of stability were done by \citet{Hayashi+22,Hayashi+23}, using N-body simulations of mildly hierarchical triples.}}

In this paper, we develop an analytical stability timescale, expanding the parameter space to \textit{non-hierarchical} systems. We consider three-body systems consisting of a primary and companion of masses $m_p$ and $m_c$, respectively, on a circular orbit, and a test particle $m_t$ $( m_p>m_c \gg m_t )$, on a highly eccentric orbit either about  the primary body or about the massive binary as a whole, such that the system is not necessarily hierarchical.  We denote these two cases the ``external companion'' and ``internal companion'' cases, respectively.
Interestingly, the stability of such systems is primarily sensitive to a single parameter, $\alpha$, defined as the ratio of the periapsis distance of the test particle to the companion's semimajor axis:
\begin{equation}\label{eq:defalpha}
    \alpha = \frac{r_{\rm peri}}{a_c} = \frac{ a_t(1-e_t) }{a_c}
    \ ,
\end{equation}
where $a_t$ and $e_t$ are the semimajor axis and eccentricity of the test particle's orbit. The system is most unstable for $\alpha=1$, when the companion orbits at the same distance as the test particle's periapsis.

We derive our stability timescale based on how long it takes for the test particle's orbital energy to change. We also show that the changes in energy are driven by the fundamental susceptibility of the eccentric test particle to energy changes at its periapsis, and not by close encounters with the companion or the primary. {Notably, we do not use the secular approximation,} in which the phases of each orbit are averaged over timescales much longer than each orbital period, {or any perturbations to the secular approximation.} In that approximation, {often used to describe the evolution of hierarchical systems}, the long-term change in the energy of each orbit is zero. {Thus, in order to explore the instability associated with orbital energy exchange, which is a feature of non-hierarchical systems,} we must describe systems by other means.

\begin{figure*}[t]
\centering
\includegraphics[width=0.98\textwidth,keepaspectratio]{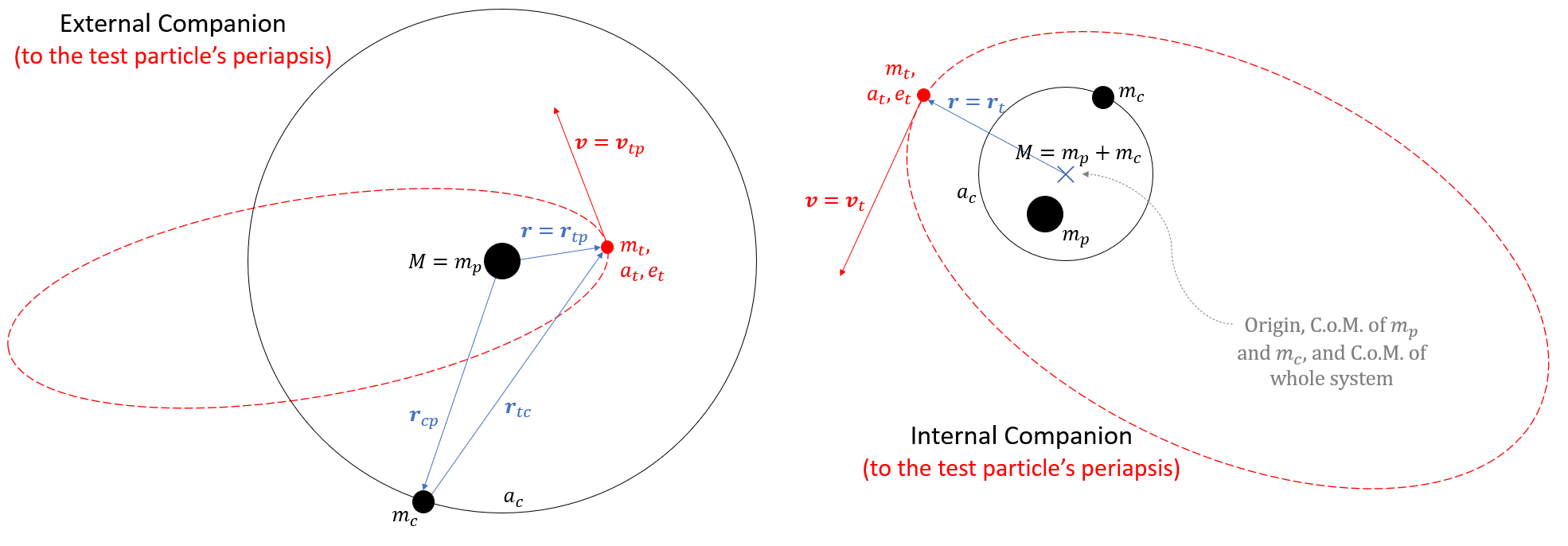} \caption{
{\bf An illustration of the possible configurations.} {\bf Left Panel:} A system where the companion orbits external to the test particle's periapsis. The central body is the primary, so that the test particle's orbital elements are defined via the position vector ${\bf r} = {\bf r}_{tp}$ and the velocity vector ${\bf v} = {\bf v}_{tp}$. {\bf Right Panel:} An internal companion. The effective central ``body'' is at the center of mass of the primary and the companion, and is fixed at the origin. The test particle's orbit is calculated using ${\bf r} = {\bf r}_t$ and ${\bf v} = {\bf v}_t$.}\label{fig:Cartoon}
\end{figure*}

In particular we are motivated by the configuration at the center of our galaxy, which includes many crossing orbits. As an example, we focus on the hypothetical existence of an intermediate mass black hole (IMBH) that may orbit around the supermassive black hole (SMBH). The existence of such an IMBH has been investigated based on a combination of theoretical and observational studies \citep[e.g.,][]{Hansen+03,Maillard+04,Grkan+20,Gualandris+09,Chen+13,Generozov+20,Fragione+20,Zheng+20,Naoz+20,Gravity+20,Rose+22}. The IMBH may have a crossing orbit with the well studied star, S0-2. This star is located close to the SMBH Sgr A*. It has an orbital period of $16$ years and an eccentricity of about $0.88$. The recent closest approach of this star has been  used to test and confirm the prediction of general relativity (GR) for the relativistic redshift  \citep[e.g.,][]{GRAVITY18,Do+19} and the advance of the periapsis 
\citep{Gravity+20}. Thus, in principle, S0-2 may be used to constrain the possibility of the existence of such an IMBH.
In addition, many of the S-star cluster orbits Sgr A* at similar distances, with many potential orbital crossings \citep[e.g.,][] {Ghez+05,Gillessen+09,Yelda+14}. One may wonder how unstable these orbits truly are, and if the S-star cluster is in fact a stable configuration.

The paper is organized as follows: we develop in Section \ref{sec:Analytic} an analytic timescale $T_{\rm stab}$ for the time it takes for the energy of the test particle to change significantly. In Section \ref{sec:Numeric} we then adopt, as a proof of concept, the possible existence of an intermediate-mass black hole (IMBH) at the center of our galaxy, which we treat numerically, and compare the results with the analytic timescale. In Section \ref{sec:App} we discuss applications of our timescale to various astrophysical settings. Finally, in Section \ref{sec:Conclusion} we summarize our results.

\section{ Analytic Stability Timescale }\label{sec:Analytic}

In this section, we present an analytic derivation of the stability timescale. For pedagogical purposes, we derive the timescale using two approaches: via the test particle's energy evolution, and via its semimajor axis evolution. The two quantities are defined by the equation,
\begin{equation}\label{eq:defEnergy}
    E = \frac{1}{2} {\bf v}^2 -
    \frac{ G M }{ r } =
    - \frac{ G M }{ 2 a_t }
    \ ,
\end{equation}
where $E$ is the (specific) energy\footnote{This definition of the energy includes only the binding energy between the test particle and the central body, and ignores the interaction energy with other bodies. This definition of the orbital energy allows the second equality, involving the osculating semimajor axis $a_t$, to hold.} of the test particle's orbit, $a_t$ is the test particle's semimajor axis, $M$ is the mass of the central body or system, $r$ is the distance between the test particle and the central body, and ${\bf v}$ is the velocity vector of the test particle relative to the central body. By this relation, the energy and semimajor axis contain the same information, so that the a change in one of the quantities by a given factor necessarily accompanies a change in the other by the same factor. 

Throughout the derivation, and the remainder of the paper, we adopt the following notation: the masses of the primary, companion, and test particle will be denoted by $m_p$, $m_c$, and $m_t$ respectively. The semimajor axis, eccentricity, inclination, longitude of ascending node, argument of periapsis, and true anomaly of both orbits will be denoted by $a$, $e$, $\iota$, $\Omega$, $\omega$, and $f$, with a subscript $c$ indicating the companion's orbital elements, and the subscript $t$ indicating the test particle's orbital elements. The companion's and test particle's orbital periods are denoted $P_c$ and $P_t$, respectively.

We fix the origin at the center of mass of all three bodies. Note that, as $m_t \rightarrow 0$, this is equivalent to the center of mass of the primary and companion. Position vectors of each of the bodies will be denoted by ${\bf r}$, and velocity vectors will be denoted by ${\bf v}$, with subscripts $p$, $c$, and $t$ for the primary, companion, and test particle. The relative position and vectors between bodies are denoted by ${\bf r}_{ij} = {\bf r}_i - {\bf r}_j$, where $i$ and $j$ are the subscripts of the bodies. The relative velocity vectors are similarly denoted by ${\bf v}_{ij}$. We will write $r$ and $v$ for the magnitudes of these vectors.

As stated in the introduction, we assume a mass hierarchy between the three bodies, $m_p > m_c \gg m_t$, with $m_t \rightarrow 0$ as it is a test particle.
We assume a highly eccentric\footnote{$e_t \gtrsim 0.5$; see Appendix A for a discussion of the low-eccentricity regime.} test particle orbit and a circular companion orbit ($e_c$ = 0).

Because we are working in the non-hierarchical regime, where the orbital distances are not well-separated, defining the orbits of the companion and the test particle can be ambiguous. In particular, the mass of the central body, $M$, defined in Eq.\ (\ref{eq:defEnergy}), depends on the configuration of the system. We define the companion's orbit to be external (internal) to the test particle's orbit if its semimajor axis\footnote{Since the companion's orbit is circular, $a_c$ is the same as the orbital distance, which is constant.}, $a_c$, is greater (less) than the test particle's periapsis distance, $r_{\rm peri} = a_t (1-e_t)$. If the companion is external, then the test particle's orbit is calculated with respect to the primary, and the central mass is $M = m_p$. If the companion is internal, then the test particle's orbit is calculated with respect to the center of mass of the primary and the companion, i.e., the origin, and the central mass is $M = m_p + m_c$. Fig.\ \ref{fig:Cartoon} shows the two possible configurations.

\subsection{ Energy Approach }

\begin{figure*}[t]
\centering
\includegraphics[width=0.47\textwidth,keepaspectratio]{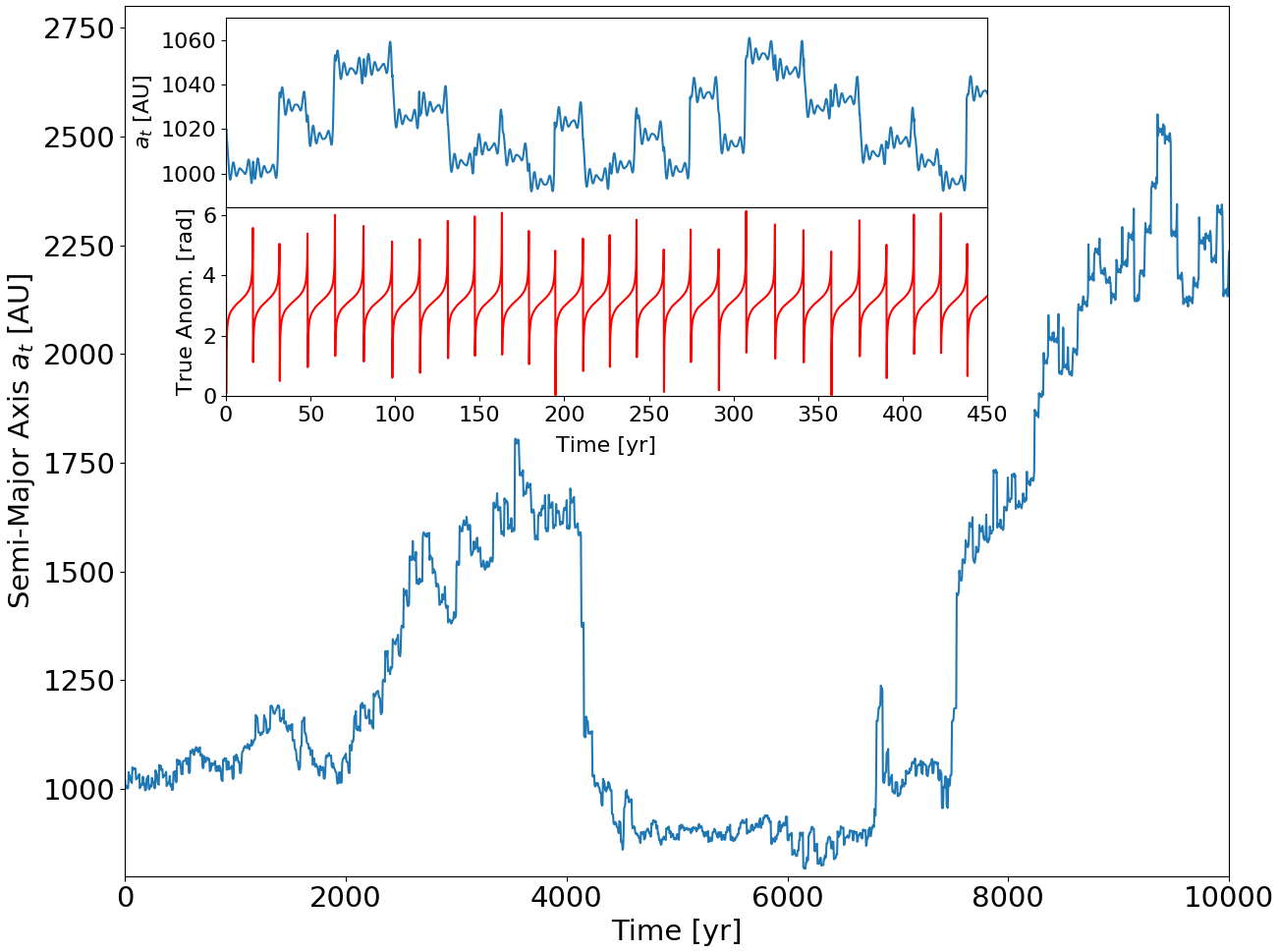}
\includegraphics[width=0.47\textwidth,keepaspectratio]{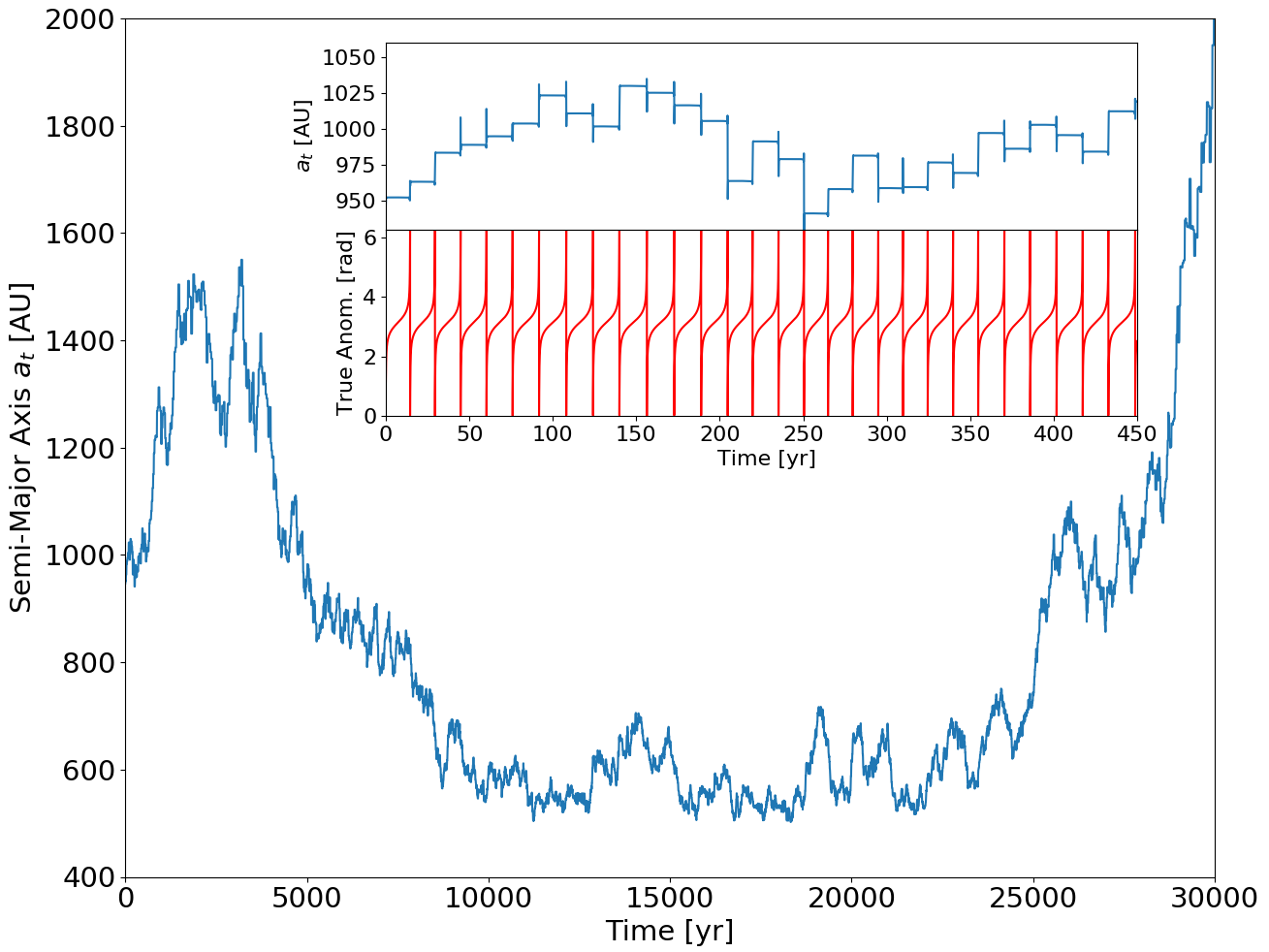}
    \caption{ {\bf Sample time evolution of the test particle's semimajor axis}. The semimajor axis evolution contains the same information as the energy evolution. {\bf Left Panel; External companion.} An external companion IMBH of 9000 $\msun$ orbiting Sgr A* at 380 au, and a test particle star orbiting Sgr A* with an initial semimajor axis of 1020 au and eccentricity of 0.884. The semimajor axis can increase and decrease with no clear bias, thus justifying the approximation of the energy evolution as a random process. We also show a ``zoomed in'' time evolution of the same system. Substantive changes to the semimajor axis of the star only occur during the ``peaks'' in the true anomaly evolution, corresponding to the periapsis passage of the star. {\bf Right Panel; Internal companion. } An internal companion IMBH of 27000 $\msun$ orbiting Sgr A* at 64 au, and a test particle star orbiting the Sgr A*-IMBH binary with an initial semimajor axis of 1020 au and eccentricity of 0.884. The system's behavior is similar to the external case.}
    \label{fig:Timeevo}
\end{figure*}

Our derivation is based on the realization that when the test particle has a highly eccentric orbit, its energy is especially susceptible to gravitational perturbations near its periapsis.
{
Accordingly, to evaluate the effect of these encounters, we approximate the energy evolution}
as a series of discrete, {\it small}, jumps, where the jumps occur at each periapsis passage. The energy may either increase or decrease at each jump, depending upon the location of the companion body at that time, 
{and so we treat its evolution as a random walk.} We justify these assumptions analytically later in this section, and numerically in Section \ref{sec:Numeric} (specifically, see Fig.\ \ref{fig:Timeevo}).

Indeed, it is possible for the test particle to undergo a large change in its energy while it is not at its periapsis, such as in the case of a close encounter between the test particle and the companion, e.g., if the test particle enters within a few Hill radii of the companion. However, for the purposes of our analytic derivation, we will ignore the effect of such events, which are arguably much rarer than periapsis passages. In principle, this assumption could cause our analytic timescale to overestimate the stability of the system, but our numerical results in Section \ref{sec:Numeric} indicate that this is, in general, not the case. 

Considering only small jumps in the energy, the energy evolves in a diffusive manner, {as previously assumed.} Though the expected overall energy change after $N$ such jumps averages to zero, the mean-square energy change is nonzero. This behavior leads to a spread in the distribution of possible energies after $N$ periapsis passages, so that for sufficiently large $N$, it is probable that the energy has changed significantly, {and thus the system has become unstable.} In particular, we say that the energy has changed significantly when the standard deviation $\sigma$ of the distribution of energy changes $\Delta E$ after $N$ periapsis passages is comparable to the initial energy itself -- in other words, when
\begin{equation}\label{eq:sigmaE}
    \sigma \left( \Delta E \right) 
    = \sqrt{N} \sigma \left( \delta E \right) 
    \sim E 
    \ ,
\end{equation}
where $\sigma (\delta E)$ denotes the standard deviation of the distribution of energy jumps $\delta E$ over a single periapsis passage.

The change in energy over any short time interval is given by
\begin{equation}\label{eq:deltaE}
    \delta E = \frac{\partial E}{\partial {\bf v}} \cdot {\delta {\bf v}}
    = {\bf v} \cdot {\bf f}_{\rm pert} \, \delta t 
    \ ,
\end{equation}
where ${\bf f}_{\rm pert}$ is the perturbing force (per unit test mass) due to the companion's presence. 
From this equation, we see that the energy change over any short time interval depends only on two quantities; the test particle's speed (or magnitude of its velocity), and the component of the perturbing force directed along the velocity vector, {$f_{\rm pert}^{\parallel}$}. For highly eccentric orbits, the speed will be much greater near periapsis than at other times in the orbit. On the other hand, the force component {$f_{\rm pert}^{\parallel}$} depends strongly on the relative position of the test particle and companion, which varies quasi-randomly so that {$f_{\rm pert}^{\parallel}$} does not peak reliably. In particular, the parallel component of the force takes on essentially random values each time the test particle approaches its periapsis.
Thus, our initial assumptions are justified; jumps primarily drive the change in energy at periapsis, and such jumps will behave like a random walk. 
The standard deviation of the distribution of energy changes is estimated by the scale of these changes, i.e.,
\begin{equation}\label{eq:sigmadeltaE}
    \sigma \left( \delta E \right) \sim
    v_{\rm peri}
    \left| {\bf f}_{\rm pert} \right|
    \delta t \ .
\end{equation}
We note that this depends only on the vector magnitudes, since the relative orientation of the vectors controls the sign and the relative strength of each energy jump, but not their overall scale.

Then Eq.\ (\ref{eq:sigmaE}) can be solved for $N$ to obtain the number of test particle orbits before a significant energy change occurs. Noting that the stability timescale $ T_{ \rm stab }$ is of order $ N P_t $, we may find $T_{ \rm stab }$ via the expression,
\begin{equation}\label{eq:TstabGeneral}
    \frac{T_{\rm stab}}{P_t} \sim
    \frac{ E^2 }{ 
    v_{\rm peri}^2
    \left| {\bf f}_{\rm pert} \right|^2
    \delta t^2
    }
    \ ,
\end{equation}
from which we may obtain $T_{ \rm stab }$ in terms of known orbital parameters by estimating each of the terms $E$, $v_{\rm peri}$, $\left| {\bf f}_{\rm pert} \right|$, and $\delta t$ for various regimes.

The energy, $E$, is given by Eq.\ (\ref{eq:defEnergy}), and $v_{\rm peri}$ can be estimated by the periapsis velocity of the test particle in a standard Keplerian orbit, given by
\begin{equation}\label{eq:vperi}
    v_{\rm peri} \approx
    \sqrt{
    \frac{GM}{a_t}
    \frac{1+e_t}{1-e_t}
    }
    \ ,
\end{equation}
where, as in Eq.\ (\ref{eq:defEnergy}), $M = m_p$ for external companions, and $M = m_p + m_c$ for internal companions.

\begin{table*}[t]\label{tab:Regimes}
\begin{center}
\begin{tabular}{c c c c c c} 
\hline
Regime & Companion Orbit & Companion Period & $M$ & $ \left| {\bf f}_{\rm pert} \right| $ & $\delta t$ \\
\hline
$r_{\rm peri}<a_c$, $T_{\rm peri} \ll P_c$ & External & Long & $m_p$ & $G m_c / a_c^2$ & $T_{\rm peri}$ \\
$r_{\rm peri}<a_c$, $T_{\rm peri} \gtrsim P_c$ & External & Short & $m_p$ & $G m_c / a_c^2$ & $P_c / 2 \pi$ \\
$r_{\rm peri}>a_c$, $T_{\rm peri} \gtrsim P_c$ & Internal & Short & $m_p + m_c$ & $G m_p m_c a_c^2 / M r^4_{\rm peri}$ & $P_c / 2 \pi$\\
\hline
\end{tabular}
\end{center}
\caption{A summary of estimates of $M$, $\left| {\bf f}_{\rm pert} \right|$, and $\delta t$ for each regime. $\left| {\bf f}_{\rm pert} \right|$ is given by Eqs.\ (\ref{eq:fextEstimate}) and (\ref{eq:fintEstimate}), and $\delta t$ is given by Eq.\ (\ref{eq:dtGeneral}). }
\end{table*}

Depending on whether the companion is external or internal, there are two possible expressions for the perturbing force, ${\bf f}_{\rm pert}$.
When the companion is external, the perturbing force is given by
\begin{equation}\label{eq:fext}
    {\bf f}_{\rm pert} =
    - \frac{ G m_c {\bf r}_{tc} }{ r_{tc}^3 }
    - \frac{ G m_c {\bf r}_{cp} }{ r_{cp}^3 } \ ,
\end{equation}
where the first term is the gravitational force on the test particle from the companion, and the second term is due to the non-inertial motion of the primary about the primary-companion mutual center of mass. Under the approximation that the majority of the encounters are weak, and noting that $a_c > r_{\rm peri}$ for external companions, the distance between the two bodies, $r_{tc}$, will typically be on the order of $a_c$. In particular, $r_{tc} \sim r_{cp} \sim a_c $, so that the magnitude of the total perturbing force can be estimated as
\begin{equation}\label{eq:fextEstimate}
    \left| {\bf f}_{\rm pert} \right| \sim
    \frac{ G m_c }{ a_c^2 }
    \ .
\end{equation}

When the companion is internal, we calculate the test particle's orbit with respect to the center of mass of the primary and companion. In this case, ${\bf f}_{\rm pert}$ is
\begin{equation}\label{eq:fint}
    {\bf f}_{\rm pert} =
    - \frac{ G m_p {\bf r}_{tp} }{ r_{tp}^3 }
    - \frac{ G m_c {\bf r}_{tc} }{ r_{tc}^3 }
    + \frac{ G (m_p + m_c) {\bf r}_t }{ r_t^3 }
    \ ,
\end{equation}
where the first two terms are the total gravitational force on the test particle, and the third term subtracts the {non-perturbing} component from the total force from the primary and companion, respectively. The magnitude of the perturbing force is best estimated by the {leading-order} multipole term of the force, which is the quadrupole force. Thus, we can approximate the magnitude of Eq.\ (\ref{eq:fint}) as\footnote{ { A more detailed derivation of this estimate can be found in Appendix \ref{sec:QuadDeriv}. } }
\begin{equation}\label{eq:fintEstimate}
    \left| {\bf f}_{\rm pert} \right| \sim
    \frac{ G m_p m_c a_c^2 }{ M r^4_{\rm peri} }
    \ .
\end{equation}

Roughly speaking, one can consider two possible cases for the interaction interval $\delta t$. The first, is the aforementioned periapsis interval, under the approximation that the most significant perturbations (though still weak) occur over some interval of time at which the test particle is near its periapsis, $T_{\rm peri}$.
{{
Since the test particle's susceptibility to energy exchange is determined by its velocity (as per Eq.\ (\ref{eq:deltaE})), we estimate the length of this interval by the time integral of the cross-track component of the relative test particle velocity, $v_{\phi}$, over one period of its orbit, divided by its maximum value, i.e.,
\begin{equation}\label{eq:Tperi}
    T_{\rm peri} =
    \int_0^{P_t} \frac{v_{\phi}}{ v_{\phi, \rm max} } \, dt
    =
    (1-e_t) P_t \ .
\end{equation}
}}

If the companion does not move significantly during the interval of periapsis passage, then the perturbing force can be treated as constant over $T_{\rm peri}$. In this case, the energy change over the whole periapsis passage is well-described by an impulse approximation, and the interaction interval $\delta t$ is the same as $T_{\rm peri}$.

\begin{table*}[t]
\begin{center}
\begin{tabular}{c c c c c c} 
 \hline
 Label & $m_c$  [$M_\odot$]& $a_c$ [au]& $r_{\rm peri}$ [au] & $e_t$ & No. Runs \\
 \hline
 {\it DiffComp} & $50-10^6$ & $10-10^4$ & $118.32$ & $0.884$ & $8308$ \\
 {\it DiffTestP} & $5000$  & $100$ & $16-4000$ & $0.2-0.997$ & $18440$ \\
 \hline
\end{tabular}
\end{center}
 \caption{The relevant numerical initial conditions. The results are shown in Figs.\ \ref{fig:Main} and \ref{fig:atet}. Note that in these Figures the square points represent an average over $10$ realizations in terms of the initial orbital angles, $\omega_{c,t}, \Omega_{c,t}, i_{c,t}$ and $f_{c,t}$. }
\label{tab:NumericRuns}
\end{table*}

On the other hand, if the companion moves during the test particle's periapsis passage, we can no longer use the impulse approximation. In particular, if the companion's period is sufficiently short, it may complete a significant fraction of its orbit while the test particle is at periapsis, so that the perturbing force ${\bf f}_{\rm pert}$ cannot be regarded as constant over the interval $T_{\rm peri}$. In this case, the impulse approximation is, strictly speaking, invalid. Nonetheless, an interaction timescale $\delta t$ can still be estimated. In this case, we evaluate the change in energy per periapsis passage by dividing the interval of periapsis passage into shorter intervals, $dt$, for which the perturbing force is constant. Then, by integrating over each of these intervals, the total energy change, $\delta E$, is 
\begin{equation}\label{eq:dEGen}
    \delta E =
    {\bf v} \cdot
    \int_{t_i}^{t_f} 
    {\bf f}_{\rm pert} \, dt
    = v_{\rm peri}
    \int_{t_i}^{t_f} 
    f_{\rm pert}^{\parallel} \, dt 
    \ ,  
\end{equation}
where $[t_i,t_f]$ is the interval of periapsis passage, and $f_{\rm pert}^{\parallel}$ is the component of the perturbing force parallel to the test particle's velocity during the interaction. 
Thanks to the companion's circular orbit, this component of the force varies roughly sinusoidally with the companion's mean anomaly, so that it may be approximated as
\begin{equation}\label{eq:dvsin}
    f_{\rm pert}^{\parallel} \sim 
    \left| {\bf f}_{\rm pert} \right|
    \cos n_c t
    \ ,
\end{equation}
where $n_c$ is the mean motion of the companion.
Then Eq.\ (\ref{eq:dEGen}) can be approximated as
\begin{eqnarray}\label{eq:dEGen2}
    \delta E & \, \sim \, &
    v_{\rm peri}
    \int_{t_i}^{t_f}
    \left| {\bf f}_{\rm pert} \right|  
    \cos n_c t \, dt
    \\
    & \, = \, &
    v_{\rm peri} n_c^{-1} \left| {\bf f}_{\rm pert} \right| 
    \sin n_c t 
    \Big|_{t_i}^{t_f}
    \sim
    v_{\rm peri} n_c^{-1} \left| {\bf f}_{\rm pert} \right|
    \ , \nonumber
\end{eqnarray}
where the sine term represents the dependence on the orientation and thus causes the spread in the distribution of possible $\delta E$. From Eq.\ (\ref{eq:dEGen2}), it follows that the interaction timescale when the companion's period is short is
\begin{equation}\label{eq:dtShort}
    \delta t \sim n_c^{-1} = \frac{P_c}{2 \pi} \ .
\end{equation}
In general, the interaction timescale is given by the shorter of the two cases,
\begin{equation}\label{eq:dtGeneral}
    \delta t \sim
    \min \left( 
    T_{\rm peri}, \, \frac{P_c}{2\pi}
    \right)
    \ ,
\end{equation}
so that the former (latter) value is used if the companion's movement is (is not) negligible over the periapsis interval.

There are three regimes to consider when evaluating Eq.\ (\ref{eq:TstabGeneral}): (1) external companion with a long period, (2) external companion with a short period, and (3) internal companion with a short period. The three possible cases are summarized in Table
\ref{tab:Regimes}. Note that Kepler's law $P_c \propto a_c^{3/2}$ forbids the case of an internal companion with a long period.

Evaluating Eq.\ (\ref{eq:TstabGeneral}) for the two external companion cases and the one internal companion case then gives
\begin{eqnarray}\label{eq:TstabTot}
    \frac{ T_{\rm stab} }{P_t} & \sim & 
    \frac{
        q^{-2}
    }{
        4 (1+e_t)
    }
     \\
    & \times & 
    \left\{ \begin{array}{lr}
    \alpha^{-4} (1-e_t)^3 / (2 \pi)^2   
    & \quad    T_{\rm peri} < P_c / 2\pi , ~ \alpha < 1\\
    \alpha^{-1} (1-e_t)^2 (1+q)
    &  \quad   T_{\rm peri} > P_c / 2\pi,~ \alpha < 1 \\
    \alpha^{7} (1-e_t)^2 (1+q)^4
    & \quad    \alpha \geq 1 \,
    \end{array} 
    \right. \ , \nonumber 
\end{eqnarray}
where $q$ is the mass ratio $m_c/m_p$ and $\alpha = a_t(1-e_t)/a_c$  (see Eq.\ (\ref{eq:defalpha})).

\subsection{Semimajor Axis Approach}

The energy of the test particle orbit is proportional to the inverse of its osculating semimajor axis; that is, $E \propto a_t^{-1}$. Thus a change in $E$ by an order of itself necessarily means that $a_t$ has changed by an order of itself as well. Similar to Eq.\ (\ref{eq:deltaE}), a significant change in the test particle's semimajor axis occurs when
\begin{equation}\label{eq:sigmaSMA}
    \sigma( \Delta a_t ) = \sqrt{N} \sigma( \delta a_t ) \sim a_t \ .
\end{equation}

The change in semimajor axis at each periapsis passage, $\delta a_t$, can be estimated by
\begin{equation}
    \delta a_t \approx
    \frac{da_t}{dt}
    \Big|_{\rm peri}
    \delta t
    \ ,    
\end{equation}
where $da_t / dt$, given by the Lagrange Planetary Equation for semimajor axis evolution (e.g., \citet{MD00book,PW2014}; where we adopt similar notation to \citet{Will21LagrangeQ}), is:
\begin{eqnarray}\label{eq:Fulldadt}
    \frac{da_t}{dt} = & \, &
    2 \sqrt{ \frac{a_t^3}{ GM (1-e_t^2) } } \times
    \\
    & \, &
    \left\{
    e_t \sin f_t \, {\bf f}_{\rm pert} \cdot {\bf \hat{i}} \,
    + (1+e_t \cos f_t) \, {\bf f}_{\rm pert} \cdot {\bf \hat{j}} \,
    \right\} \ , \nonumber 
\end{eqnarray}
where $\bf{\hat{i}}$ and $\bf{\hat{j}}$ are the radial and cross-track unit vectors of the test particle with respect to the central body.

Given the interaction timescale mentioned above (Eqs. (\ref{eq:Tperi}) and (\ref{eq:dtShort})), $\delta a_t$ is, at periapsis,
\begin{equation}
    \delta a_t \approx
    2 \sqrt{ \frac{a_t^3 (1+e_t)}{GM (1-e_t) } } 
    \, {\bf f}_{\rm pert} \cdot {\bf \hat{j}} \, \delta t
    \ .
\end{equation}
Information about the relative orientation of the two orbits is contained by the ${\bf f}_{\rm pert} \cdot {\bf \hat{j}}$ term. The standard deviation of the distributions of possible $\delta a_t$ is then on the order of
\begin{equation}
    \sigma (\delta a_t) \sim
    2 \sqrt{ \frac{a_t^3 (1+e_t)}{ GM (1-e_t) } } \left| {\bf f}_{\rm pert} \right| \delta t \ ,
\end{equation}
which, when plugged into Eq.\ (\ref{eq:sigmaSMA}), for the various cases of $\left| {\bf f}_{\rm pert} \right|$ and $\delta t$, yields the same timescale as Eq.\ (\ref{eq:TstabTot}).

\section{Comparison with Numerical Results}\label{sec:Numeric}

\subsection{Description of the Initial Conditions and the Numerical Approach}\label{sec:ICs}

As a proof of concept, we consider a system consisting of the galactic center black hole Sgr A* as the primary mass $m_p$, a star in orbit around Sgr A* as the test particle $m_t$, and a hypothetical IMBH companion to Sgr A* as the companion $m_c$. Using \texttt{HNBody} \citep[][]{Rauch+02}, we directly integrate over 26,000 variations of this system assuming Newtonian gravity\footnote{We do not include any post-Newtonian (PN) precession for the purposes of this comparison. The 1st pN precession can stabilize the system against secular perturbations \citep[e.g.,][]{Naoz+12GR}  even in the cases where the orbits are more compact than the hierarchical limit \citep[e.g.,][]{Wei+21,Faridani+21}. However, the purpose of this numerical analysis is to test our analytical timescale.}. 
From the results, we may calculate $T_{\rm stab}$ numerically by finding the first time at which the test particle's semimajor axis (energy) changes by a factor of 2 from its initial value. Comparing this value to the analytic prediction then provides a test of the analytic timescale. {We note that not every {\it destabilization}, as defined above, necessarily leads to an ejection of the particle from the system. Ejection occurs when the energy changes from negative to positive (i.e., $\Delta E > +E$), and is thus a subset of destabilization.}

\begin{figure*}[t]
\centering
\includegraphics[width=0.99\textwidth,keepaspectratio]{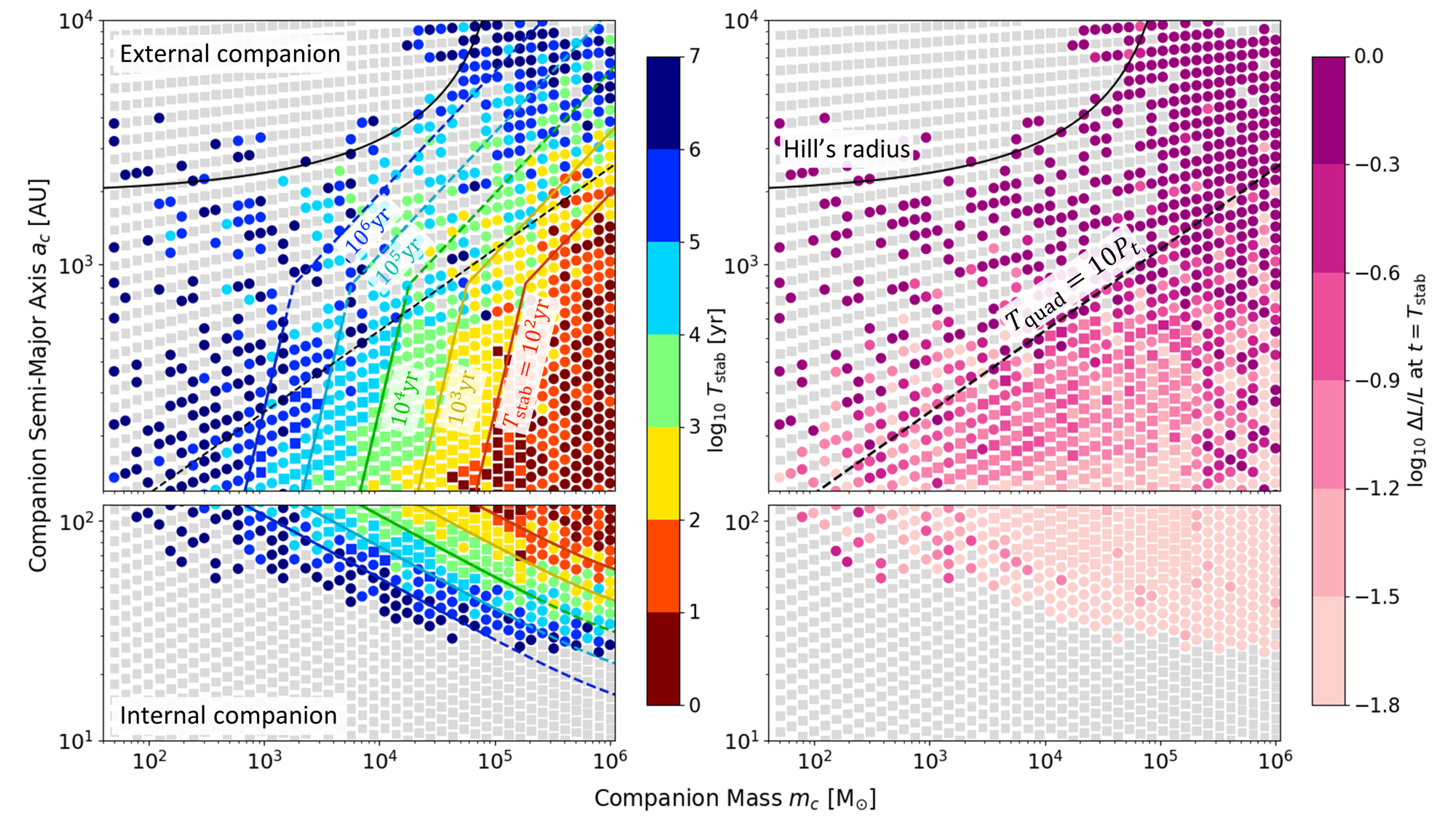}
\caption{ { \bf Stability timescale as a function of companion parameters.}
{\bf Left}: $T_{\rm stab}$ as a function of the companion's semimajor axis $a_c$ and its mass $m_c$, as determined by direct N-body integration. The upper (lower) panels describe an external (internal) companion. A colored dot represents a change in S0-2's energy by a factor of more than 2 within the maximum integration time of $10^7$ yr, where the color determines $T_{\rm stab}$. A colored square represents 10 systems with the same $m_c$ and $a_c$, such that all 10 systems have a change in the test particle energy of more than a factor of 2 within the integration time. The color of the square is the geometric mean of $T_{\rm stab}$ over all 10 systems. A gray square represents no such change in the test particle energy within the integration time. Overlaid are lines of constant  $T_{\rm stab}$, as predicted by Eq.\ (\ref{eq:TstabTot}). Some systems outside the regime where $T_{\rm stab} < 10^7$ yr still exhibit significant change in their orbit, if they are within the region where the test particle can enter the Hill sphere of the companion, i.e., below the solid black line. The dashed black line marks where the Kozai quadrupole timescale is ten times greater than the test particle period. Secular approximations are reasonably applicable for systems above this line, causing them to exhibit angular momentum oscillations which can suppress the random walk nature of the energy evolution, leading to more stable systems than predicted by the analytic timescale.
{\bf Right}: The normalized change in test particle angular momentum at $t=T_{\rm stab}$. As predicted by the quadrupole timescale, systems above this line exhibit significant changes in angular momentum.
}
\label{fig:Main}
\end{figure*}

We run a grid of systems, varying the IMBH mass and separation. In all runs, we use $m_p = 4 \times 10^6 ~\msun$ and $m_t = 10~\msun$, and set the companion's eccentricity to $e_c = 10^{-3} \approx 0$.\footnote{ {When $e_c=0$ exactly, the argument of periapsis $\omega_c$ is ill-defined. Eccentricities of exactly zero are unlikely in practice, so we avoid this pathological case by setting $e_c=10^{-3}$.} } We present two sets of numerical runs, as summarized in Table \ref{tab:NumericRuns}. In the runs labeled {\it DiffComp}, we set the test particle's initial orbital parameters to be the same in all runs, namely those of the star S0-2, i.e., $a_t=1020~{\rm au} \,,~e_t=0.884$, and $m_t=10$~M$_\odot$, and vary the companion's mass and semimajor axis systematically. The test particle's argument of periapsis $\omega_t$, longitude of ascending node, $\Omega_t$, and inclination $\iota_t$, are chosen such that they will have the observed values of S0-2 on the sky, and then projected into the invariable plane. 
The companion for this set of runs is varied on a grid of $m_c\in[50,10^6]$~M$_\odot$, and $a_c\in[10,10^4]$~au. 

In the second set of runs (labeled {\it DiffTestP}), we set the companion's mass and semimajor axis to be $m_c=5000$~M$_\odot$ and $a_c=100$~au, respectively, and its eccentricity to be $e_c = 10^{-3}$. In this case, we arbitrarily choose the inclination of the companion's orbit to be close to zero\footnote{ {When $\iota_c=0$ exactly, the longitude of ascending node $\Omega_c$ is ill-defined.} } ($0.001$~rad). This choice eliminates dependencies on the companion's argument of periapsis and longitude of ascending nodes. 
In this set of runs, the test particle's initial periapsis distance is then systematically varied from $a_t (1-e_t)\in[16,4000]$, and its eccentricity from $e_t\in[0.2,0.997]$.

When the orbital configurations are varied, in runs {\it DiffComp} or {\it DiffTestP}, we chose the initial longitudes of ascending nodes, arguments of periapsis, and mean anomalies from a uniform distribution between $0$ and $2\pi$, and the mutual inclination from an isotropic distribution (i.e., uniform in cosine).

In Section \ref{sec:NumTimeEvo} we discuss the numerical time evolution of the test particle orbit, and how it justifies the assumptions used in deriving the analytic timescale. In Sections \ref{sec:DiffComp} and \ref{sec:DiffTestP} we separately analyze each set of numerical runs and compare them to the analytic timescale predictions.

\subsection{Time Evolution of the Test Particle Orbit}\label{sec:NumTimeEvo}

The numerical results justify our approximation of the test particle's energy evolution as a series of discrete, random ``jumps'' occurring each time the test particle reaches its periapsis. Fig.\ \ref{fig:Timeevo} depicts two representative cases of the test particle's time evolution in the presence of an external and internal companion (left and right panels, respectively). In particular, the inset panels show a zoom-in on the part of the evolution that exhibits the energy jumps described in Section \ref{sec:Analytic}. These jumps occur at the test particle's periapsis, as indicated by the peaks in its true anomaly evolution. Moreover, Fig.\ \ref{fig:Timeevo} demonstrates that over time the test particle's semimajor axis (energy) can increase and decrease in a diffusive manner so that its evolution is well described by a random process.
We provide additional examples of the time evolution of the test particle's orbit in the Appendix (see Fig.\ \ref{fig:detailed}).

\hspace{0.5cm} \hfill \\

\subsection{Dependence of the Timescale on the Companion Parameters}\label{sec:DiffComp}

In the {\it DiffComp} set of runs, we integrated a total of $8308$ configurations, initially starting with the same test particle orbital configuration and varying systematically the orbit of the companion. All integrations are stopped at  $10^7$ years.

As stated above, we record the time at which the test particle's energy has changed from its initial value by a factor of $2$. The results are shown in the left hand side of Fig.\ \ref{fig:Main}, where the color coding shows the time at which the test particle's energy changed by a factor of $2$, ranging from $1$ to $10^7$ years. Systems whose test particle energy never changes by a factor of $2$ within $10^7$ yr are depicted as a gray square. We depict this value as a function of the companion's semimajor axis $a_c$ and mass $m_c$ (filled circles in Fig.\ \ref{fig:Main}). For each point on the grid, we initially run one system with a value of $m_c$ and $a_c$. The top panel shows systems with an external companion ($a_c>r_{\rm peri}$), and the bottom panel shows systems with an internal companion ($a_c<r_{\rm peri}$).

On the right hand side of Fig.\ \ref{fig:Main}, we plot the fractional change in angular momentum (i.e., $\Delta L/L$) at $t=T_{\rm stab}$. The symbols have the same meaning as the left hand side panel.

Note that in the part of the parameter space where the system is very non-hierarchical ($a_c \sim r_{\rm peri})$ and where $m_c \ll m_p$, the dynamical behaviour is rather chaotic \citep[e.g.,][]{Stone+19,Ginat+21,Ginat+21X,Kol+21,Manwadkar+21}. Therefore, in this part of the parameter space we ran an additional 10 configurations with the same values of $m_c$ and $a_c$, and numerically calculated $T_{\rm stab}$ for these values by taking the geometric mean of the times at which the energy changes by a factor of $2$. These grid points are marked as boxes in Fig.\ \ref{fig:Main}. Over-plotted are the contours of the analytic timescales from Eq.\ (\ref{eq:TstabTot}). 

For the majority of the parameter space, in particular where the system is ``least hierarchical'' ($\alpha \sim 1$), we find good agreement between the analytic and numeric results. This suggests both that our analytic result is a good estimate of the stability timescale, and also that the instability is sufficiently explained by energy jumps at periapsis. Close encounters of the test particle with the other two bodies, while possible in principle, are evidently not frequent enough to drive the instability of the system in this regime.

There are two regions of the parameter space for which there is disagreement between the analytic and numerical results: systems above the black dashed line in Fig.\ \ref{fig:Main}, which are more stable than predicted, and some systems to the left of the blue $T_{\rm stab}=10^6$~yr line, which are less stable than expected.

In the former region, the semimajor axes of the two orbits are not well-separated, but the secular approximation is nonetheless somewhat applicable. Thus, the semimajor axis evolution of these systems cannot be treated as random walks, as in Fig.\ \ref{fig:Timeevo}, {{
but oscillations that are roughly consistent with secular behavior (i.e., quasi-secular).
}}

\begin{figure}[h]
\centering
\includegraphics[width=0.47\textwidth,keepaspectratio]{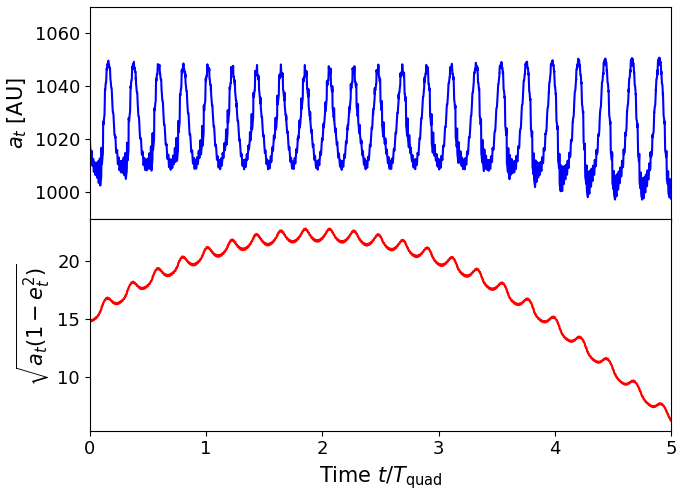}
\caption{ 
\textbf{ Time evolution of a sufficiently hierarchical system. } A system consisting of an external IMBH with mass $9000$ $\msun$ orbiting Sgr A* at $a_c = 980$ au, and a test particle star with $a_t = 1020$ au, $e_t = 0.884$. Although the system is not strictly hierarchical, the system exhibits behavior predicted by the secular approximation. The test particle's semimajor axis does not evolve like a random walk, but instead oscillates. Large eccentric Kozai-Lidov oscillations in the test particle's angular momentum occur as well. Because the system does not behave like a random walk, the analytic timescale fails.
}
\label{fig:Secular}
\end{figure}

As an example, we show the evolution of such a system in Fig.\ \ref{fig:Secular}. In this system, the energy (i.e., the semimajor axis) evolves not as a random walk, but as an oscillation, never straying far from its initial value. \footnote{
{{
These energy oscillations are similar to those noted by \citet{Antognini+14, Luo+16, Grishin+17, Grishin+18QuasiSec, Bhaskar+21}, which take place roughly on an orbital timescale and may eventually cause non-secular evolution.
}}} As expected, thanks to the Eccentric Kozai-Lidov mechanism \citep[EKL][]{Kozai,Lidov,Naoz+11sec}, the system also exhibits large oscillations in the test particle's angular momentum, shown in the upper region of the top right panel of Fig.\ \ref{fig:Main}. 
These angular momentum oscillations take place roughly on the timescale of
\begin{equation}\label{eq:tEKL}
    T_{\rm quad} \sim \frac{16}{30\pi} \frac{m_p+m_c + m_t}{m_c}\frac{P_c^2}{P_t}(1-e_c^2)^{3/2} \ , 
\end{equation}
\citep[e.g.,][]{Antognini15}. {{ Systems for which }} this timescale is much longer than the test particle's period, i.e., by a factor of 10 
{{($T_{\rm quad} > 10 P_t$), are typically governed by quasi-secular rather than random walk behavior. Thus, their stability is better described by secular criteria \citep[e.g.,][]{Ivanov+05,LN11,Katz+12,NaozSilk14,Antonini+14,Bode+14,Antognini+14, Luo+16, Grishin+18QuasiSec, Bhaskar+21}. On the upper right panel of Fig. \ref{fig:Main}, these systems are shown to have changes in angular momentum $\gtrsim$ $50\%$ and lie above the black dashed line.}}

For sufficiently large $\alpha$ (bottom panel, internal companion), the system is also hierarchical, so that secular approximations can likewise be applied, and the energy does not behave like a random walk. Here we expect that the secular inverse Eccentric Kozai-Lidov mechanism \citep[iEKL][]{Naoz+17,Naoz+20} {{or the hierarchical stability timescale of \citet{Mushkin+20}}} provides a better stability criterion.

In the latter region, some systems undergo changes in their orbital energy on shorter timescales than predicted by our analytical criterion. In such cases, close encounters between the test particle and another body are important to the system's evolution. Such close encounters may only occur if the star's apoapsis is within a few Hill radii of the IMBH's orbit. This condition can be expressed as \begin{equation}\label{eq:Hill}
    a_c < a_t (1+e_t) + k R_{\rm Hill} \ ,
\end{equation}
where $R_{\rm Hill}$ is the Hill radius of the companion, defined as 
\begin{equation}\label{eq:HillDef}
  R_{\rm Hill} = a_c \left(\frac{m_c}{m_p}\right)^{1/3} \ ,
\end{equation}
and $k$ is a factor of order unity \citep[similarly to planet-planet scattering, e.g.,][]{Chatterjee+08}. In Fig.\ \ref{fig:Main} this condition is shown as the solid black curve (where we adopt $k=3$). Indeed, systems above this curve very rarely exhibit a large change in energy. While such systems in which close encounters are important do exist, they are relatively few and far between in Fig.\ \ref{fig:Main}, whereas the general trend in Fig.\ \ref{fig:Main} obeys the timescale as predicted by energy jumps at periapsis. This provides further evidence that energy jumps at periapsis, not close encounters, are the primary drivers of instability.

\begin{figure}[t]
\centering
\includegraphics[width=0.47\textwidth,keepaspectratio]{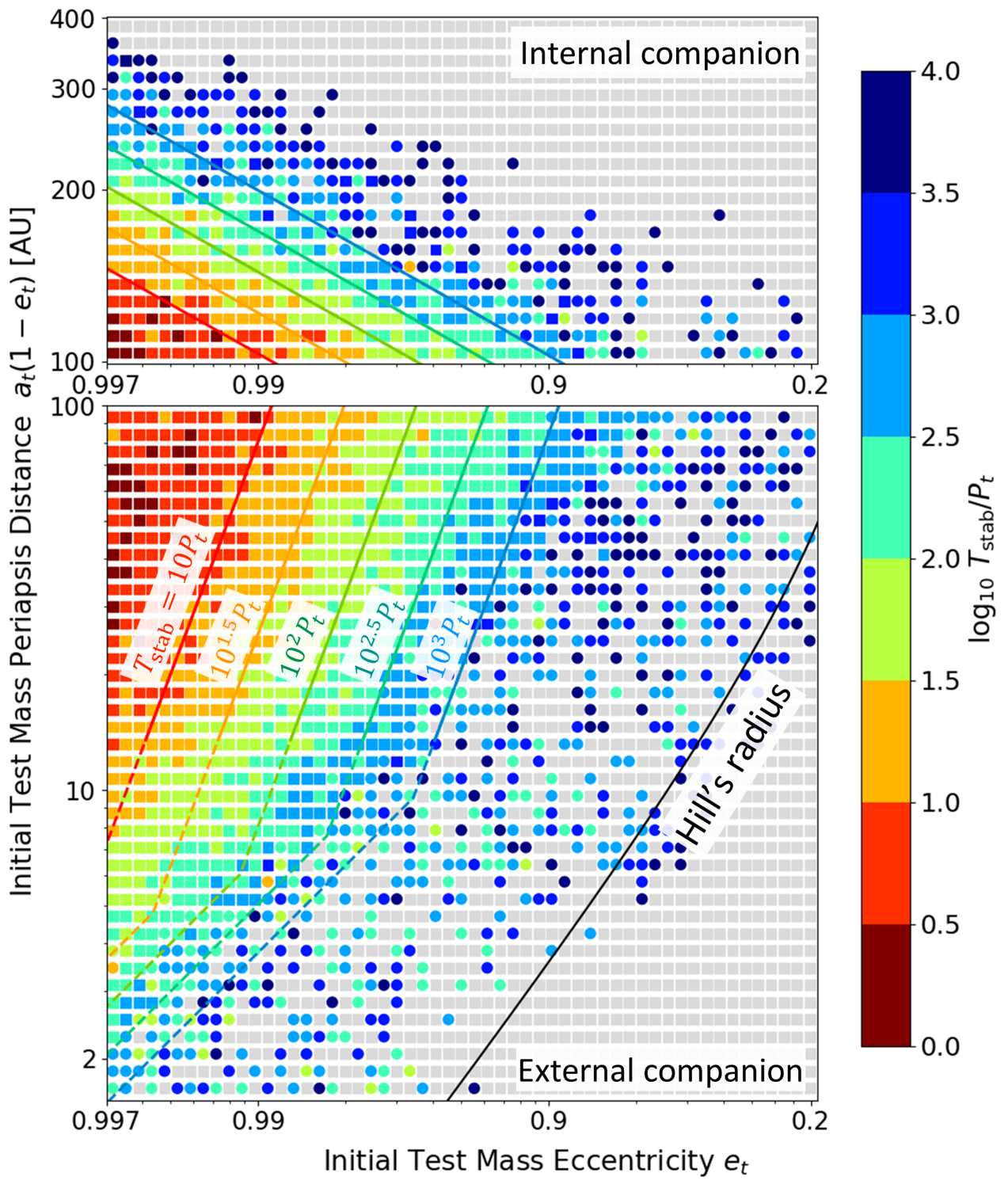}
    \caption{ { \bf Stability timescale as a function of test particle parameters.} Note that unlike in Fig.\ \ref{fig:Main}, we vary the test-particle's orbital parameters, not the companion's parameters, and thus the positions of the external and internal cases are flipped. The meaning of the symbols is the same as in Fig.\ \ref{fig:Main}. The solid black line shows the Hill sphere boundary; very few systems below this line exhibit a significant energy change within the maximum integration time of $10^4$ test particle periods.}
    \label{fig:atet}
\end{figure}

\subsection{Dependence of the Timescale on the Test-Particle Parameters}\label{sec:DiffTestP}

In this set of runs, {\it DiffTestP}, we integrated a total of $18440$ configurations, this time starting with the same companion parameters (see Table \ref{tab:NumericRuns}) and systematically varying the parameters of the test-particle. We again record the first time at which the test particle's energy changes by a factor of $2$, and depict this time as a function of the initial test particle periapsis distance $r_{\rm peri}$ and its eccentricity $e_t$. We integrated each system for $10^4 P_t$, where $P_t$ is the initial test particle period. Similarly, as for the {\it DiffComp} run, we run additional $10$ realizations for systems that have $\alpha \sim 1$. The results are shown in Fig.\ \ref{fig:atet}. The color code here represents the time for changing the energy by an order of itself, normalized by the test particle period, i.e., $T_{\rm stab}/P_t$.  In Fig.\ \ref{fig:atet} the internal case is shown in the top panel and the external case is shown in the bottom panel. 

We again find good agreement between the analytic and numerical results. As in \textit{DiffComp}, we find that the analytic and numerical timescales agree for a majority of the parameter space. Similarly to the \textit{DiffComp} set of runs, the agreement with the analytic timescale is best in the regime $\alpha \sim 1$, while the analytic timescale begins to fail in the regimes $\alpha \gg 1$ or $\alpha \ll 1$, i.e. when the system is hierarchical. Additionally, we see systems to the right of the blue $T_{\rm stab} = 10^3 \, P_t$ line which become unstable much faster than the analytic timescale predicts, which we attribute to close encounters between the companion and the test particle.

\section{Applications}\label{sec:App}

\subsection{The S-cluster in the presence of an IMBH}

Recent gravitational-wave observations by the LIGO-Virgo collaboration have now confirmed the existence of IMBHs \citep[e.g., GW190521;][]{GW190521a+20,GW190521b+20}. Specifically, our galactic center may harbor IMBHs as the result of a possible minor merger with a low-mass or dwarf galaxy, or even with a globular cluster. Such a scenario was considered by \citet{RM13}, who suggested that if IMBHs serve as the seeds of SMBHs in the center of galaxies, hierarchical galaxy evolution would yield many IMBHs in our galaxy. Additionally, a combination of theoretical and observational arguments suggest that an IMBH is expected to exist in the central parsec of our galaxy \citep[e.g.,][]{Hansen+03,Maillard+04,Grkan+20,Gualandris+09,Chen+13,Generozov+20,Fragione+20,Zheng+20,Naoz+20}. Of particular interest are the proposed efficient formation {{of IMBHs}} as a result of black-hole mergers \citep[e.g.,][]{Fragione+21}, or via black-holes collisions with main sequence stars \citep[e.g.,][]{Rose+22}.

To constrain the parameter space of mass and semimajor axis of a hypothetical IMBH , \citet{Naoz+20} used the long baseline of observations of the star S0-2, located close to the SMBH Sgr A*.
S0-2 has been observed for more than two decades, and its orbit is sufficiently regular that, if there is a companion to  Sgr A*, it is either quite close to the main black hole, or well outside the orbit of S0-2.

\begin{figure}[h]
\centering
%\vspace{-0.3cm}
\includegraphics[width=0.45\textwidth,keepaspectratio]{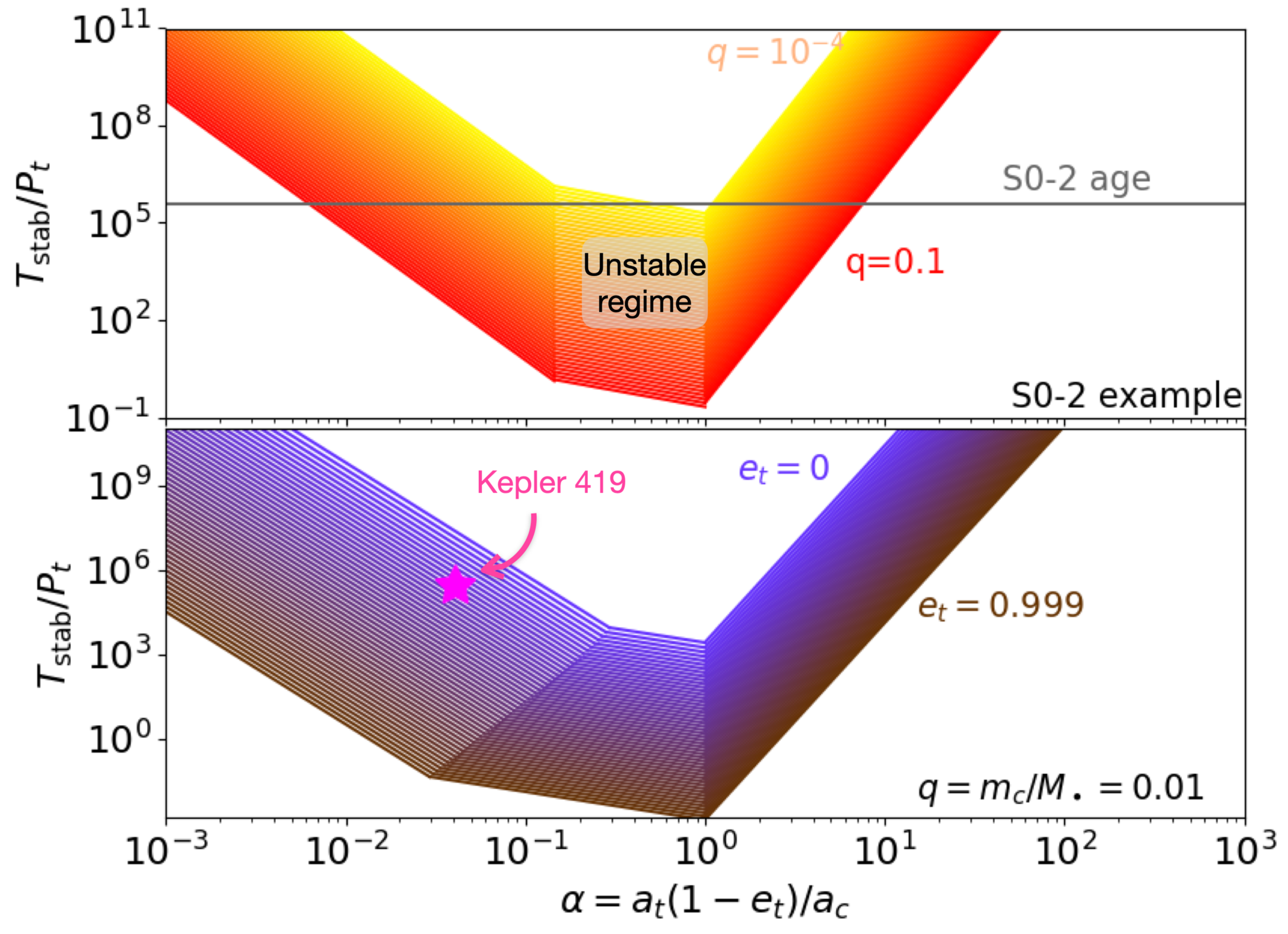}
\caption{
{\bf Stability Parameter.} {\bf Top:} Number of stable orbits of S0-2 as a function of $\alpha$, for companion-primary mass ratios between $10^{-4}$ and $10^{-1}$, assuming an eccentricity of 0.884 for S0-2. The gray line shows the age of S0-2. Regions where the gray line lies above a chosen colored line (corresponding to a given companion-primary mass ratio) are excluded values for $\alpha$, as in these cases S0-2's orbit would be too unstable to survive to its current age. {\bf Bottom:} Number of stable orbits of a test particle with the same semimajor axis as S0-2 as a function of $\alpha$ for test particle eccentricities ranging from $0$ to $0.999$, assuming a companion-primary mass ratio of 0.01 ($m_c = 4 \times 10^4 ~ M_\odot$).  The star in the bottom panel represents the approximate position of Kepler-419 (see text for discussion).
}
\label{fig:Stabparam}
\end{figure}

The stability analysis done here (see Fig.\ \ref{fig:Main}) suggests that the parameter space for which the orbit of S0-2 will be stable over its lifetime {{ \citep[6 Myr,][]{Lu+13} }} is constrained to the left side of Fig.\ \ref{fig:Main}.  Another way to visualise the parameter space is depicted in Fig.\ \ref{fig:Stabparam}. In the top panel of Fig.\ \ref{fig:Stabparam}, we consider S0-2 parameters for a wide range of a companion semimajor axis and mass ratio (with respect to the SMBH's mass), depicted as the color-coded lines. The y-axis represents the stability timescale normalized to the period of S0-2. As shown in the Figure, the stable regime for a given mass ratio $q$ is wherever the corresponding colored line lies above the grey line denoting the age of S0-2.  Thus, {{for example,}} for $q=0.1$ ($m_c = 4 \times 10^5 ~M_\odot$), the allowed regions are $\alpha \lesssim 10^{-2}$ (companion IMBH well outside the periapsis of S0-2), and $\alpha \gtrsim 10$ (companion well inside the periapsis of S0-2).  For $q = 10^{-4}$ ($m_c =400 ~ M_\odot$) (very low-mass companion), the orbit of S0-2 is stable for essentially any value of $\alpha$.

\subsection{The stability of the S-star cluster}

The S-star cluster is a collection of stars on nearly isotropic orbits, within the innermost arcsecond of Sgr A* \citep[e.g.,][]{Lu+09,Sabha+12,Gillessen+12}. These stars are eccentric and orbit-crossing, and are estimated to be relatively young \citep[e.g.,][]{Ghez+04,ghez:2005dq,Lu+09,Lu+13,Gillessen+09,Gillessen+12,Yelda+14,gillessen:2017aa,Do+19}. Thus, in the context of this investigation the question of the stability of the S-cluster translates to the timescale of the stability. We simplify the question by considering perturbations to S0-2's orbit by other stars, one at a time. 

Examining the top panel of Fig.\  \ref{fig:Stabparam} we see that the low mass ratio curve ($q=10^{-4}$) implies that even an object as massive as $400$~M$_\odot$ will be able to lurk in the presence of S0-2 without destabilizing its orbit during its lifetime. Thus, other stars, naturally will not destabilize S0-2's orbit as well. We thus suggest that during the lifetime of the S-cluster, star-star interactions do not change the stellar orbits' energies by order of themselves\footnote{Note that other processes such as collisions and tidal capture also take place on longer timescales  than the S-cluster lifetime \citep[e.g.,][]{Rose+20,Rose+22}, thus the conclusion here can be extended to other physical processes beyond two-body interactions.}. 

\subsection{Stellar and Planetary Systems}

Since the stability timescale developed here depends only on products of ratios and other unitless parameters, it is not limited to galactic nuclei and may be applied in a wide range of systems.
Consider, for example, a planetary system. Interestingly, the forces between the S-stars and the SMBH are similar to those in a planetary system with a Sun-like star in the presence of a companion. In particular, the mass ratio between Sgr A* and the stars in the S-cluster is similar to that of a star and its planetary system. Further, the distance scales at the galactic center (SMBH and S-Star, and SMBH-companion) are roughly the square of those for a star-planet, and a star-companion.
Thus, our stability timescale is also relevant in the case of planetary systems.

Significantly, observations suggest that most of the massive stars reside in binaries or higher multiples \citep[e.g.,][]{Raghavan+06,Raghavan+10,Sana+12,Moe+17,Moe+21}.  In the planetary field one also often considers orbital crossing as a sign of instability \citep[e.g.,][]{Sourav+08,Nag+11,Denham+19,Wei+21,Faridani+21}. However, we suggest that the stability question needs to be cast as the {\it timescale} to instability in exoplanetary systems as well.

As an example, consider the bottom panel of Fig.\ \ref{fig:Stabparam}. This Figure depicts a system with mass ratio of $q=0.01$ between the primary and the companion, and varies the test-particle eccentricity.  Consider a system of a star and a brown dwarf companion on a circular orbit. It is thus easy to see that a circular planet (the test-particle), or even a planet with a non-negligible eccentricity on a crossing orbit, will not have its energy changed by a factor of itself for a timescale between $2000$ and $10^9$ orbital periods of the planet, depending on the value of $\alpha$.
This implies that some planetary systems may even be observed near instability.  Of course, a smaller companion results in an even longer stability timescale. 

A potentially relevant example may reside in the observed Kepler-419 system. This system has two massive Jupiters orbiting a $1.4 \,M_\odot$ star,  with $m_t = 2.5 \,M_J$, $e_t = 0.83$,  $a_t = 0.37$ au, and $m_c = 7.3 \,M_J$, $e_c = 0.18$, $a_c= 1.7$ au, respectively. \citep[e.g.][]{Dawson+12,Dawson+14}. The system's stability has been questioned recently by \citet{Denham+19} and \citet{Jackson+19}. For this system, $P_t = 0.19$ yr, $P_c = 1.84$ yr, $T_{\rm peri} =0.11 (P_c/2\pi)$, and our hierarchical parameter  is $\alpha=0.037$. Thus, 
we may ask on what timescale the system's energy will change by an order of itself.
Using the first line of Eq.\ (\ref{eq:TstabTot}), we obtain $T_{\rm stab} \sim 7 \times 10^4$ yr.
This is shown as the red star on the bottom panel of Fig.\ \ref{fig:Stabparam}. Therefore, at face value, this suggests that a major change in the inner planet's energy could occur within fewer than $10^5$ years.

We note that the inner planet cannot really be considered as a test particle in this system. However, our numerical investigation was done for a non-negligible mass inner star and is roughly consistent with the analytical criterion (for comparable masses, our criterion underestimates the stability timescale). Thus, our stability timescale for Kepler-419 may not be far from reality.
  
\section{Conclusion}\label{sec:Conclusion} 
  The long-term stability of three-body systems is a fundamental problem in astrophysics with many applications, from clusters of stars in galactic nuclei to planetary systems. Furthermore, the instability of triple systems is associated with energy exchange between the two orbits, and is a time-sensitive concept. In this paper, we develop an analytic timescale at which instability sets in for {{non-hierarchical triple systems.}} We find that numerical N-body experiments corroborate this analytic result. In summary:
  \begin{enumerate}
      \item
      The timescale for which the system is stable, $T_{\rm stab}$, is given by Eq.\ (\ref{eq:TstabTot}), assuming a circular companion and highly eccentric test particle orbit {{
      in a non-hierarchical configuration. This analytic result is validated numerically in Figs.\ \ref{fig:Main} and \ref{fig:atet}.
      }}
      
      \item {{For systems with such non-hierarchical configurations, the}} instability is primarily driven by the high susceptibility to energy change the test particle experiences near its periapsis, rather than by close encounters between the companion and the test particle. 
      The evolution of the test particle's orbital energy behaves like a random walk, {{ as in Fig.\ \ref{fig:Timeevo}. }}

      \item {{
      In the hierarchical limit, the random walk assumption on the energy evolution does not hold, and so hierarchical systems are stable for much longer than predicted by Eq.\ (\ref{eq:TstabTot}). In particular, their stability is predicted by secular evolution. Systems do not need to be strictly hierarchical for the random walk assumption to fail; they only need to be sufficiently well-described by the secular approximation, such as in Fig.\ \ref{fig:Secular}.}} Large oscillations in the angular momentum caused by the Eccentric Kozai-Lidov effect are an indicator of secular-like behavior.
      
  \end{enumerate}
  
\begin{acknowledgments}
  {EZ thanks Evgeni Grishin for helpful comments and discussion.} SN acknowledges partial support from NASA ATP  80NSSC20K0505 and   thanks Howard and Astrid Preston for their generous support. CMW is grateful for the hospitality of the Institut d'Astrophysique de Paris, where parts of this work were carried out, and acknowledges partial support from the National Science Foundation, Grant Nos.\ PHY 19-09247 and 22-07681.
\end{acknowledgments}

% B stands for Bigwhale!

\appendix

\section{The Stability Timescale at Lower Eccentricities}\label{sec:lowecc}

\begin{figure}[h]
\centering
\includegraphics[width=\textwidth,keepaspectratio]{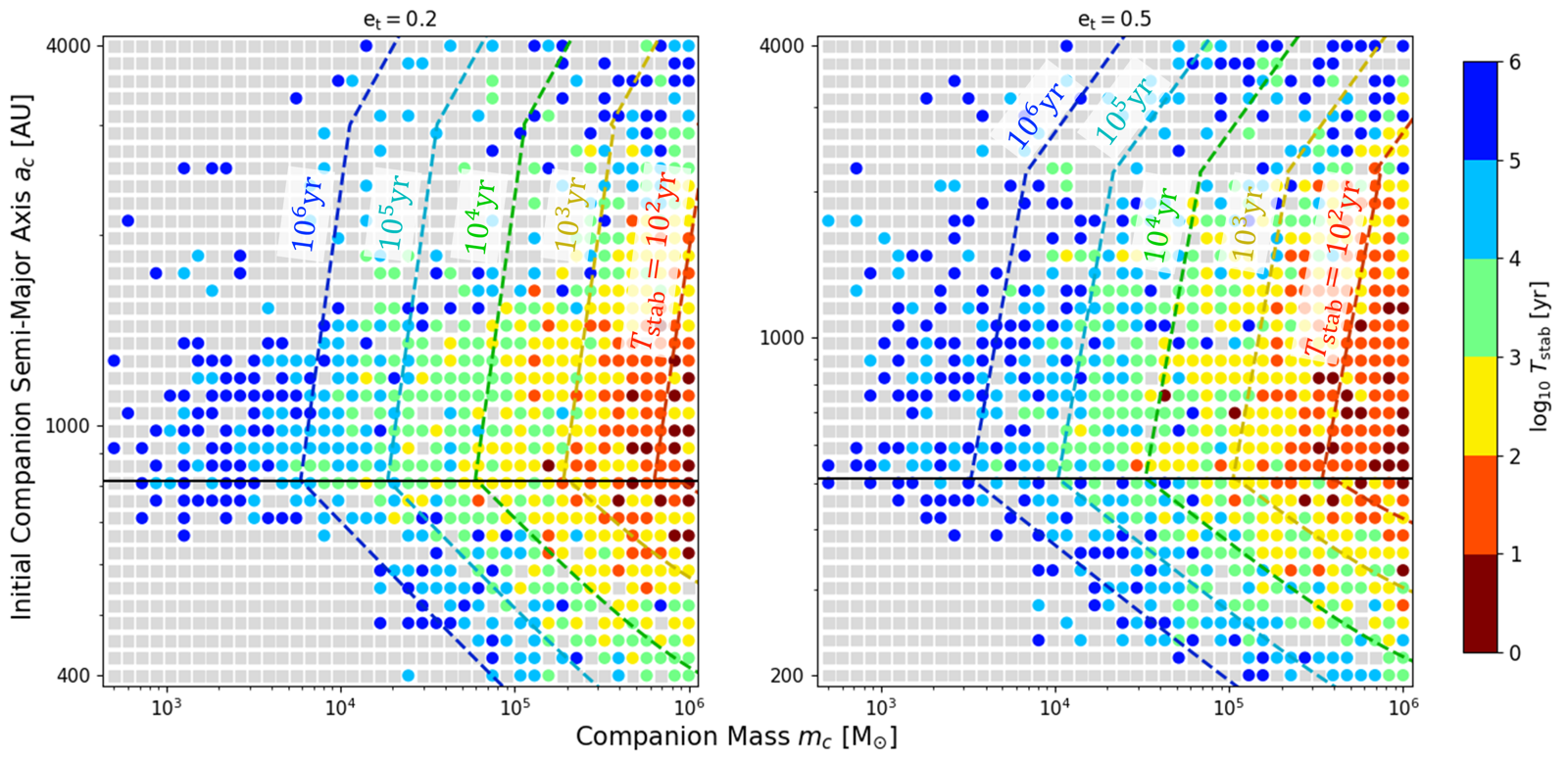}
\caption{
{\bf Stability Timescale for Low Eccentricities.} $T_{\rm stab}$ as a function of $m_c$ and $a_c$ as determined by direct N-body integration, for a test particle star of semimajor axis 1020 au, and with eccentricities 0.2 and 0.5. The analytic timescale overestimates the numerical timescale.
}
\label{fig:rparam}
\end{figure}

Our analytic calculation relies on the assumption that significant changes to the test particle's energy occur near its periapsis. For this assumption to hold, it is necessary to assume a highly eccentric test particle orbit, because of the dependence of $dE/dt$ on the test particle's velocity, which is sharply peaked about the periapsis only for eccentric orbits. When these assumptions are violated, the analytic timescale does not strictly hold. Nevertheless, we show numerically that the analytic timescale still has some applicability at lower eccentricities.

To study the behavior of systems with low test particle eccentricity, we integrate two sets of 1554 systems, with fixed test particle orbital configuration and systematically varied companion orbital parameters. In one set of systems, the test particle has an initial eccentricity of 0.2, and the initial companion semimajor axis varies from 400 to 4000 au. In the other set, the test particle has an initial eccentricity of 0.5, and the initial companion semimajor axis varies from 200 to 4000 au. In both sets, the initial test particle semimajor axis is 1020 au, the companion mass is varied from $500$ to $10^6$ $\msun$, and the initial orbital angles of both the test particle and companion are randomly drawn uniformly in $\Omega, \omega, f$ from $[0,2\pi]$ and uniformly in $\cos i$ from $[-1,1]$. All systems are integrated for $10^6$ yr.

Our numerical results show that, for lower eccentricities, the analytic timescale {{overestimates}} the numerically calculated timescale in regions of the parameter space where orbital crossing is possible. Thus, the energy jumps at periapsis, as described in the main body of the paper, cannot fully explain the instability in the low eccentricity regime. For a test particle eccentricity $e_t = 0.5$, this discrepancy is not major. For $e_t$ as low as $0.2$, the disagreement becomes more severe, with the analytic result overestimating the numeric timescale by roughly a full order of magnitude.

\section{Supplemental Equations and Plots}

We present supplemental equations and plots to our analytic derivation in Section \ref{sec:Analytic}. This section will assume familiarity with the logical framework of Section \ref{sec:Analytic}, but will provide additional rigor and justification for several arguments made in the section. In particular, we will derive Eq.\ (\ref{eq:deltaE}), the energy jump per periapsis passage, and Eqs.\ (\ref{eq:fext}) and (\ref{eq:fint}), the perturbing force vectors, directly from the Newtonian equations of motion, and provide additional detail for our estimate of $ \left| {\bf f}_{\rm pert} \right| $ for internal companions in Eq.\ (\ref{eq:fintEstimate}). {{We also provide an additional, more detailed, description of the time evolution of the systems studied in the paper.}}

\subsection{ The Perturbing Force }

The time derivative of $E$, as defined by Eq.\ (\ref{eq:defEnergy}), is, by direct differentiation,
\begin{equation}\label{eq:dEdt}
    \frac{dE}{dt} =
    {\bf v} \cdot
    \left[
        \frac{ d^2 {\bf r} }
        { dt^2 }
        -
        \frac{ GM {\bf r} }
        { r^3 }
    \right] \ ,
\end{equation}
where ${\bf r}$ and ${\bf v}$ are the position and velocity vectors of the test particle relative to the central body, and $r$ is the magnitude of ${\bf r}$.

The first term in brackets is the relative acceleration of the test particle, and the second term is equivalent to the acceleration of the test particle due to the central body. Thus the entire quantity in brackets, which is the difference of the two, is the component of the test particle acceleration that is not due to the central body. We call this term the perturbing force (per test mass), i.e.,
\begin{equation}
    {\bf f}_{\rm pert} = 
    \frac{ d^2 {\bf r} }
    { dt^2 }
    -
    \frac{ GM {\bf r} }
    { r^3 } \ .
\end{equation}
When the perturbing force is defined this way, Eq.\ (\ref{eq:dEdt}) is equivalent to Eq.\ (\ref{eq:deltaE}).

Eqs.\ (\ref{eq:fext}) and (\ref{eq:fint}) then follow immediately from the Newtonian equations of motion for ${\bf r}$. For an external companion, ${\bf r} = {\bf r}_{tp}$, and the corresponding equation of motion is \begin{equation}\label{eq:EoMExt}
    \frac{d^2 {\bf r}_{tp} }{ dt^2 } =
    - \frac{ Gm_c {\bf r}_{tc} }{ r_{tc}^3 }
    - \frac{ Gm_p {\bf r}_{tp} }{ r_{tp}^3 }
    - \frac{ Gm_c {\bf r}_{cp} }{ r_{cp}^3 } \ .
\end{equation}
For an internal companion, ${\bf r} = {\bf r}_t$, and
\begin{equation}\label{eq:EOMInt}
    \frac{ d^2 {\bf r}_t }{ dt^2 } =
    - \frac{ Gm_p {\bf r }_{tp} }{ r_{tp}^3 }
    - \frac{ Gm_c {\bf r }_{tc} }{ r_{tc}^3 } \ .
\end{equation}

\begin{figure}[t]
\centering
\includegraphics[width=0.99\textwidth,keepaspectratio]{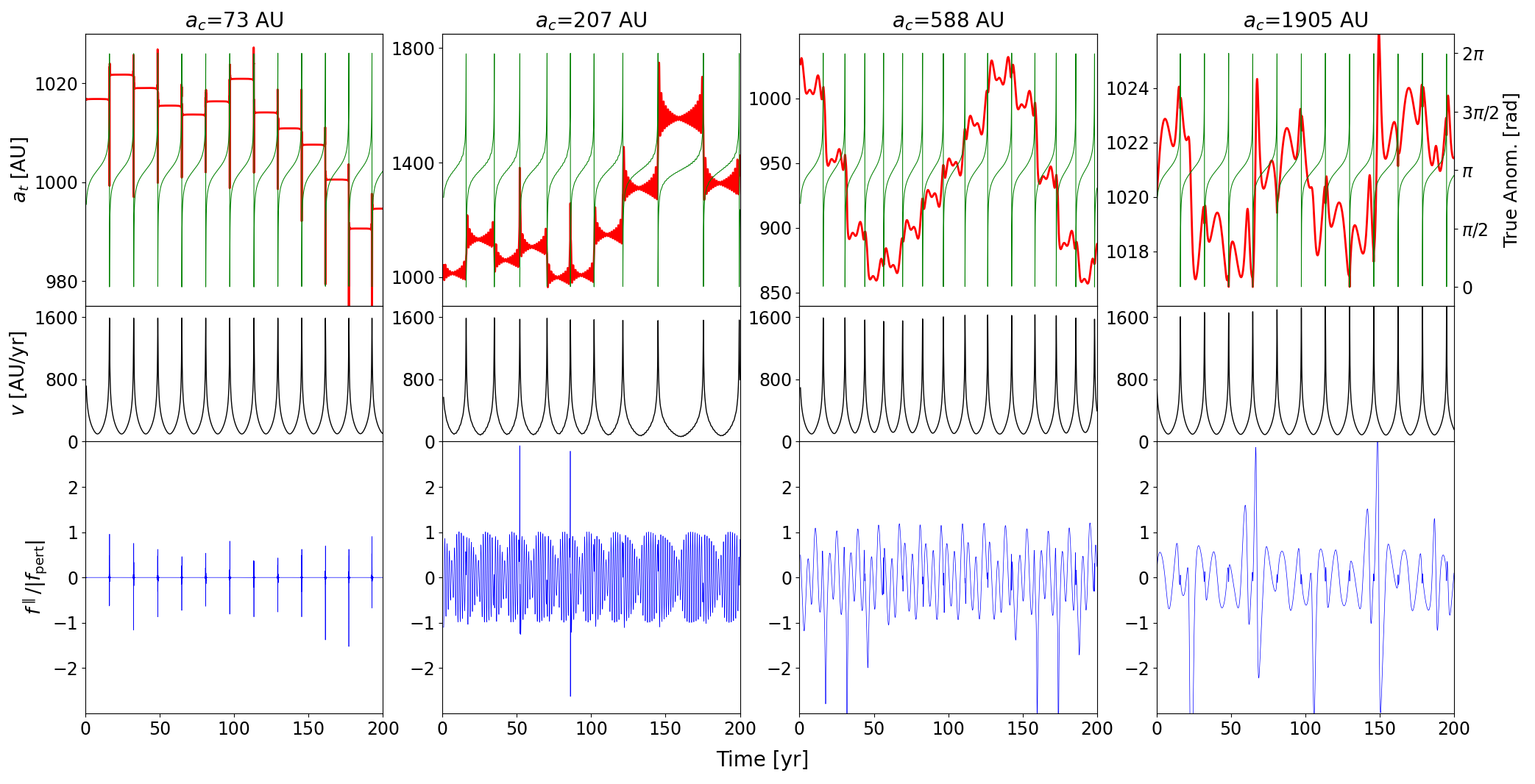}
\caption{ 
\textbf{Additional example time evolutions of the test particle's orbit.} Each column shows the evolution of a system similar to those in the \textit{DiffComp} set of runs. For all systems $m_c = 27000 \, \msun$, while $a_c$ varies between each system. \textbf{Top Row:} The evolution of the test particle's semimajor axis $a_t$ (red) and true anomaly $f_t$ (green). \textbf{Middle Row:} The magnitude of the test particle's relative velocity, $v$. \textbf{Bottom Row:} The ratio between $f^{\parallel}$, the component of ${\bf f}_{\rm pert}$ parallel to the relative test particle velocity $\bf v$, and $| {\bf f}_{\rm pert} |$, given by Eqs.\ (\ref{eq:fextEstimate}) and (\ref{eq:fintEstimate}).
}
\label{fig:detailed}
\end{figure}

\subsection{ Estimating $\left | {\bf f}_{\rm pert} \right|$ for Internal Companions }\label{sec:QuadDeriv}

When the companion is internal, the gravitational potential experienced by the test particle can be expanded in a multipole series, by
\begin{equation}
    \Phi =
    - \frac{ Gm_p }{ \left| {\bf r} - {\bf r}_p \right| }
    - \frac{ Gm_c }{ \left| {\bf r} - {\bf r}_c \right| }
    =
    - \sum_{l=0}^{\infty}
    \frac{ G a_c^l \eta_l }{ r^{ \, l+1} }
    P_l ( \cos \theta )
    \ ,
\end{equation}
where $P_l$ are the Legendre polynomials, $\theta$ is the angle between the companion and the test particle's position vectors, and
\begin{equation}
    \eta_l =
    \frac{
    m_p (-m_c)^l + m_c m_p^l
    }
    { (m_p + m_c)^l }
    \ .
\end{equation}
The $l=0$ term is the potential due to the central body, so the perturbing force is the negative spatial gradient of all remaining terms, which is
\begin{equation}
    {\bf f}_{\rm pert} =
    \sum_{l=2}^{\infty}
    {\bm \nabla}
    \left[
        \frac{ G a_c^l \eta_l }{ r^{ \, l+1} }
        P_l ( \cos \theta )
    \right]
    = 
    \sum_{l=2}^{\infty}
    \frac{ G a_c^l \eta_l }{ r^{ \, l+2} }
    {\bm \xi}_l
    \ ,
\end{equation}
where ${\bm \xi}_l$ is a vector, with magnitude of order unity, specifying the direction of the force, given by
\begin{equation}
    {\bm \xi}_l =
    P'_l \idotI
    \left[
    \hat{\bf I} - \idotI \hat{\bf i}
    \right]
    - (l+1) P_l \idotI \hat{\bf i}
    \ ,
\end{equation}
where $\hat{\bf I}$ is the unit vector of the companion's position, and $\hat{\bf i}$ is the unit vector of the test particle's position. 

The leading term of the remaining series is the $l=2$ or quadrupole term, since $\eta_1 = 0$. Then $\left | {\bf f}_{\rm pert} \right|$ is well estimated by the leading order coefficient, $G a_c^2 \eta_2 / r^4 $, evaluated at $r = r_{\rm peri}$, yielding Eq.\ (\ref{eq:fintEstimate}). This estimate is only good to an order of magnitude; the higher order terms, while decreasing, are not negligible since the ratio in which the series is expanded in, $a_c/r$, is less than unity but not small. Thus, the higher order terms contribute nontrivially to the total force, but not at a larger scale than the leading order.

\subsection{ The Time Evolution  }

Fig.\ \ref{fig:detailed} shows additional examples of the time evolution of the test particle's orbit, for various regimes. 

As was shown in Section \ref{sec:NumTimeEvo}, in all regimes, significant changes in the test particle's semimajor axis occur when its velocity peaks; that is, in a short interval near its periapsis. However, as discussed in Section \ref{sec:DiffComp}, such jumps need not behave like a random walk. The two right columns in Fig.\ \ref{fig:detailed} show systems in the quasi-secular regime. In the first system, changes in the semimajor axis are dominated by periapsis jumps, but these jumps behave in an oscillatory manner. In the second system, the jumps are small as the companion is distant from the test particle's periapsis, so that the semimajor axis evolution is dominated by quasi-secular effects.

The bottom panels of each column indicate that the ratio between $f^{\parallel}$, the component of ${\bf f}_{\rm pert}$ parallel to the relative test particle velocity $\bf v$, and our estimate $\left| {\bf f}_{\rm pert} \right|$ of the perturbing force, given by Eqs.\ (\ref{eq:fextEstimate}) and (\ref{eq:fintEstimate}), is of order unity. This indicates that $\left| {\bf f}_{\rm pert} \right|$ is indeed a good order-of-magnitude estimate of the perturbing force. Note that for external companions, the estimate is valid at all points in the test particle's orbit, but for internal companions (represented by the leftmost panel in Fig.\ \ref{fig:detailed}), the estimate is only valid near periapsis, since it was derived with that assumption.

\bibliographystyle{aasjournal}
\bibliography{Binary}
\end{document}